\newcommand{\bra}[1]{\langle #1|}
\newcommand{\ket}[1]{|#1\rangle}
\documentclass{PoS}

\title{Form factors for semi-leptonic B decays}

\ShortTitle{Form factors for semi-leptonic B decays}

\author{Ran Zhou, \speaker{Steven Gottlieb}%
	\\
        Department of Physics, Indiana University, Bloomington, IN 47405, USA\\
        E-mail: \email{sg@indiana.edu}}

\author{Jon A.~Bailey\\
        Department of Physics and Astronomy, Seoul National University, Seoul 151-747, South Korea\\
	}

\author{Daping Du, Aida X.~El-Khadra, R.D.~Jain\\
        Physics Department, University of Illinois, Urbana, Illinois 61801, USA\\
	}

\author{Andreas S.~Kronfeld, Ruth S.~Van de Water\\
        Fermi National Accelerator Laboratory, Batavia IL 60510, USA\\
	}

\author{Yuzhi Liu, Yannick Meurice\\
	Department of Physics and Astronomy, University of Iowa, Iowa City, IA 52240, USA\\
	}
\author{(Fermilab Lattice and MILC Collaborations)}

\abstract{
We report on form factors for the $B\rightarrow K  l^+ l^-$ 
semi-leptonic decay process.  
We use several lattice spacings from $a=0.12$ fm down to 0.06 fm and a
variety of dynamical quark masses with 2+1 flavors of asqtad quarks
provided by the MILC Collaboration.  These ensembles allow good control
of the chiral and continuum extrapolations.
The $b$-quark is treated as a clover quark with the Fermilab interpretation.
We update our results for $f_\parallel$ and $f_\perp$, or, equivalently,
$f_+$ and $f_0$.  In addition, we present new results for the tensor
form factor $f_T$.  Model independent results are obtained based upon
the $z$-expansion.}

\FullConference{The 30th International Symposium on Lattice Field Theory\\
                 June 24--29,  2012\\
                 Cairns, Australia}

\begin{document}

\section{Motivation and theoretical background}
The rare $B$ meson decay $B \to Kll$ is mediated by a flavor
changing neutral current (FCNC).  The Standard Model (SM) contribution
occurs through a penguin diagram for the $b\rightarrow s l l$ process.
Since FCNCs are small in the SM, this decay presents an opportunity to 
detect beyond-the-Standard-Model (BSM) physics.  This decay has been studied by
several experiments including BaBar~\cite{Aubert:2003cm}, 
Belle~\cite{Abe:2001dh}, CDF~\cite{Aaltonen:2011cn}, and
LHCb~\cite{Aaij:2012cq}.
We expect that LHCb will continue to improve its precision, and new 
$B$ factories such as SuperB and SuperKEKB will greatly add to 
our understanding of this decay.

Recently, the CDF Collaboration published results for the 
$B^+ \to K^+ \mu^+\mu^-$ differential branching ratio
\cite{Aaltonen:2011cn}.
They compared their measurement with a decade-old prediction of the
decay form factors based on light cone sum rule (LCSR) 
calculations~\cite{Aaij:2012cq}.
There is considerable uncertainty in the LCSR form factors at small
momentum transfer $q^2$.  
Unless the uncertainty in the form factors is
reduced, our ability to search for BSM physics will be severely limited
by lack of knowledge of the SM decay rate. 

Lattice QCD enables first-principles calculations of the form factors with
controlled errors that are systematically improvable.
In fact, the first study of this decay, using
the quenched approximation, was done over 15 years ago~\cite{Abada:1995fa}.  Since then, there
have been three additional studies using the quenched 
approximation \cite{DelDebbio:1997kr,Becirevic:2006nm,AlHaydari:2009zr}.
Recently, there have been two groups studying this decay using
$2+1$ flavor improved staggered quark configurations from the MILC
collaboration~\cite{Liu:2011raa,Zhou:2011be}.



\section{Ensembles}
Table \ref{tab:ensembles} shows the 
ensembles that were used in the current calculation.
The MILC ensembles~\cite{Bernard:2001av,Aubin:2004wf,Bazavov:2009bb}
were generated using three flavors of dynamical
quarks.  The quarks are asqtad-improved staggered quarks and a Symanzik
improved gauge action is used.  The common mass (in lattice units)
of the two light sea quarks is denoted $a m_l$ and that of the strange
sea quark is $a m_s$.  We use four coarse ($a\approx 0.12$ fm), five
fine (($a\approx 0.09$ fm) and two super-fine ($a\approx 0.06$ fm)
ensembles, in order to control the chiral and continuum extrapolations.
The ratio of light-to-strange sea-quark masses varies from 0.4 to 0.05.
On each configuration, we use four source times to increase statistics.

\begin{table}[h]
 \centering
 \begin{tabular}{ccccc}
  \hline
  $\approx a({\rm fm})$ & size & $am_l/am_s$ & $N_{\rm meas}$ \\
  \hline
  0.12 & $20^3\times 64$ & 0.02/0.05 & 2052  \\
  0.12 & $20^3\times 64$ & 0.01/0.05 & 2259  \\
  0.12 & $20^3\times 64$ & 0.007/0.05 & 2110 \\
  0.12 & $20^3\times 64$ & 0.005/0.05 & 2099 \\
  \hline
  0.09 & $28^3\times 96$ & 0.0124/0.031 & 1996 \\
  0.09 & $28^3\times 96$ & 0.0062/0.031 & 1931 \\
  0.09 & $32^3\times 96$ & 0.00465/0.031 & 984\\
  0.09 & $40^3\times 96$ & 0.0031/0.031 & 1015  \\
  0.09 & $64^3\times 96$ & 0.00155/0.031 & 791 \\
  \hline
  0.06 & $48^3\times 144$ & 0.0036/0.018 & 673  \\
  0.06 & $64^3\times 144$ & 0.0018/0.018 & 827  \\
  \hline
 \end{tabular}
 \caption{Ensembles of asqtad $N_f=2+1$ configurations analyzed.}
 \label{tab:ensembles}
\end{table}

\section{Form factors}
Semileptonic decays of a heavy-light pseudoscalar meson to a pseudoscalar
or vector meson are characterized by form factors that describe matrix
elements of a hadronic current between initial and final states. 
We
concentrate here on the decay $B \to Kll$ for which we need two matrix
elements  
$\bra{K(p_K)} i\bar{s}\gamma^\mu b \ket{B(p_B)}$
and
$\bra{K(p_K)}i\bar{s}\sigma^{\mu\nu} b\ket{B(p_B)}$, where 
the $B$ meson momentum is $p_B$ and that of the kaon is $p_K$.
For the vector current
matrix element, we define two form factors $f_+$ and $f_0$,  
while for the tensor
operator, we only need a single form factor $f_T$:

\begin{eqnarray}
\bra{K} i\bar{s}\gamma^\mu b \ket{B}&=&f_+(q^2)\left(p_B^\mu+p_K^\mu-\frac{M_B^2-M_K^2}{q^2}q^\mu \right)
+f_0(q^2)\frac{M_B^2-M_K^2}{q^2}q^\mu, \label{eq:def.f+f0}\\
\bra{K}i\bar{s}\sigma^{\mu\nu} b\ket{B}&=&\frac{2f_T(q^2)}{M_B+M_K} (p_B^\mu p_K^\nu-p_B^\nu p_K^\mu), \label{eq:def.fT}
\end{eqnarray}
In these equations, $q^\mu=p_B^\mu-p_K^\mu$ is the momentum transfer, 
and $q^2$ is the outgoing dilepton invariant mass squared.
We can alternatively describe the vector current form factors
by $f_\parallel$ and $f_\perp$.  We define
\begin{eqnarray}
\bra{K} i\bar{s}\gamma^\mu b \ket{B}&=&\sqrt{2M_B}\left [ v^\mu f_\parallel(E_K)+p_\perp^\mu f_\perp(E_K)\right ],
\end{eqnarray}
where $v^\mu=p_B^\mu/M_B$ is the four-velocity of the
$B$ meson and $p_\perp^\mu=p_K^\mu-(p_K\cdot v)v^\mu$.

In lattice QCD, it is convenient to work in the rest frame of the $B$ meson.
The new form factors are considered to be functions of the kaon energy $E_K$:
\begin{eqnarray}
f_\parallel(E_K)&=&\frac{\bra{K} i\bar{s} \gamma^0 b \ket{B}}{\sqrt{2M_B}},\\
f_\perp(E_K)&=&\frac{\bra{K} i\bar{s}\gamma^i b \ket{B}}{\sqrt{2M_B}p^i_K},\\
f_T(E_K)&=&\frac{M_B+M_K}{\sqrt{2M_B}}\frac{\bra{K(k)} ib\sigma^{0i} s\ket{B(p)}}{\sqrt{2M_B}p_K^i} ,
\end{eqnarray}
where there is no sum on the index $i$ in the equations for $f_\perp$ and $f_T$.
In the $B$-meson rest frame, $q^2=(p_B-p_K)^2=M_B^2+M_K^2-2 M_B E_K$, 
so small $E_K$ corresponds to large $q^2$.

\section{Chiral-continuum extrapolation}
\begin{figure}
\begin{center}
\begin{tabular}{c c}
\includegraphics[height=0.42\textwidth]{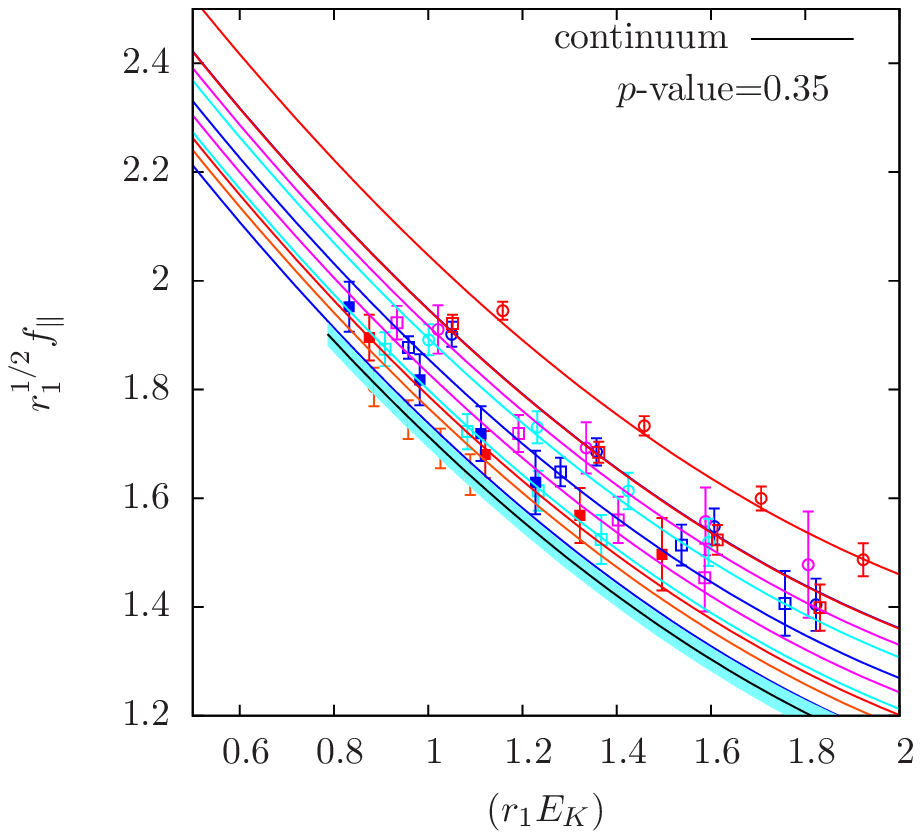}
&
\includegraphics[height=0.42\textwidth]{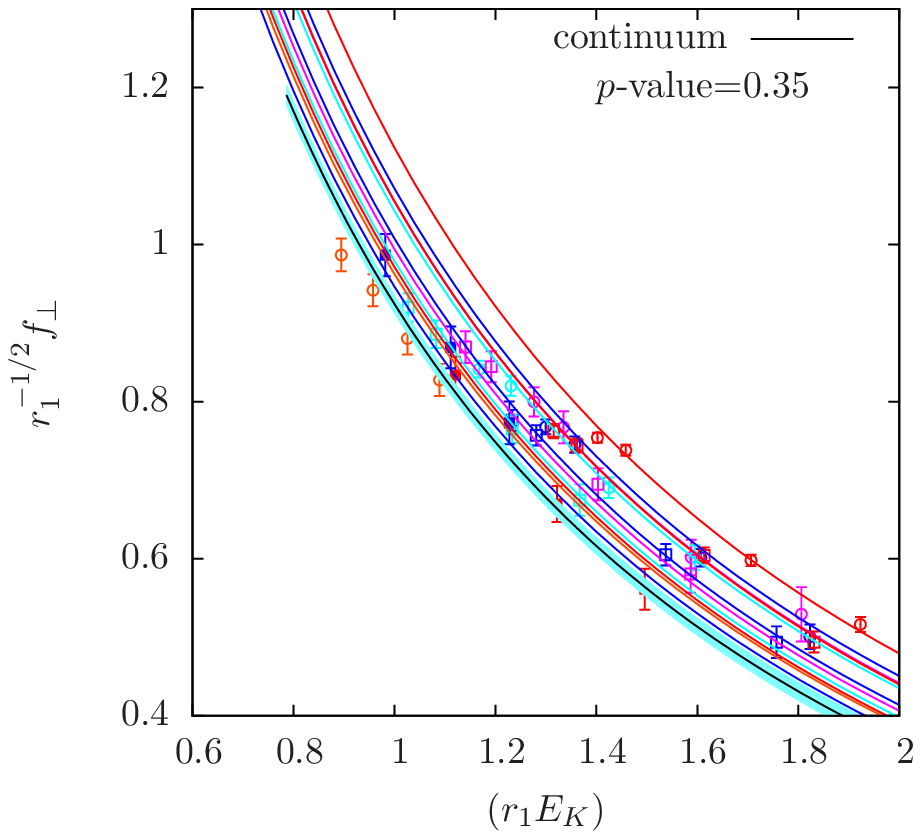}
\end{tabular}
\end{center}
\caption{The form factors $f_\parallel$ (left) and $f_\perp$ (right) as a function
of $E_K$ in lattice units.  
The legend in Fig.~2 explains the
meaning of all the symbols.}
\label{fig:fpara}
\end{figure}

\begin{figure}
\begin{center}
\includegraphics[height=0.42\textwidth]{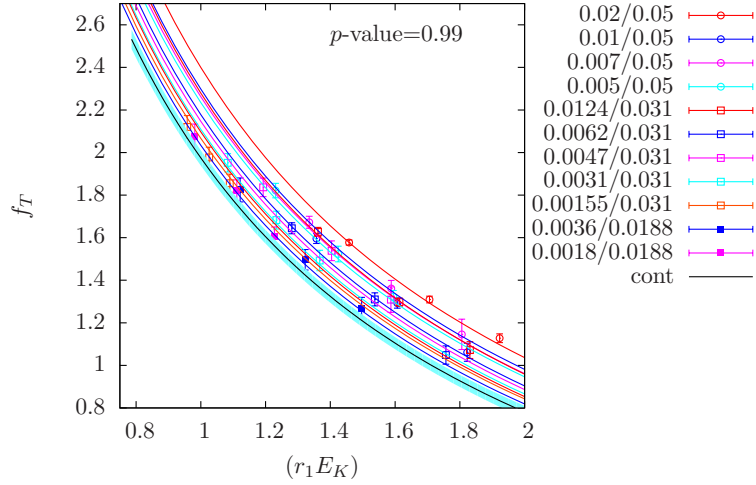}
\end{center}

\caption{The form factor $f_T$ as a function of $E_K$ in lattice units.
\label{fig:ft}
}
\end{figure}

Because we work with light quark masses larger than in Nature and at non-zero
lattice spacing, we must take the chiral and continuum limits of our
results.  We use heavy-light meson staggered chiral perturbation
theory (HMS$\chi$PT)~\cite{Aubin:2007mc} to perform extrapolations.
We work to zeroth order in $1/m_b$ and next-to-leading order in the light-quark
masses, kaon recoil energy, and lattice spacing.
The form factors $f_\parallel$ and $f_\perp$ 
are fit to the following forms:
\begin{eqnarray}
f_\parallel&=&\frac{C_\parallel^{(0)}}{f_\pi}\left [ 1+{\rm logs}+C^{(1)}_\parallel m_l+
C^{(2)}_\parallel(2m_l+m_s)+C^{(3)}_\parallel E_K+C^{(4)}_\parallel E_K^2+C^{(5)}_\parallel a^2
\right] ,\\
f_\perp&=& \frac{C^{(0)}_\perp}{f_\pi} \left [ \frac{1}{E_K+\Delta^*_{B_s}+{\rm logs}}+
\frac{\rm logs}{E_K+\Delta^*_{B_s}}\right ] \nonumber \\ &&+\frac{C^{(0)}_\perp}{f_\pi (E_K+\Delta^*_{B_s})}
\left [C^{(1)}_\perp m_l+C^{(2)}_\perp(2m_l+m_s)+C^{(3)}_\perp E_K+C^{(4)}_\perp E_K^2+ C^{(5)}_\perp a^2\right ] ,
\label{eq:chiral.fperp}
\end{eqnarray}
where the notation in Ref.~\cite{Bailey:2008wp} is used, and
$\Delta^*_{B_s}=m_{B_s^*}-M_B$.  
The 1-loop $SU(2)$ chiral logarithms (denoted
as ``logs'' above) are non-analytic functions of the pseudoscalar 
meson masses.
We use the same form for $f_T$ as for $f_\perp$, 
because the Isgur-Wise relation~\cite{Isgur:1990kf} requires that $f_T=f_+$
in the large $M_B$ limit, and the
shape of $f_+$ is dominated by $f_\perp$ at low recoil.
Figures \ref{fig:fpara} and \ref{fig:ft} show our results 
for $f_\parallel$, $f_\perp$ and $f_T$ including the continuum, physical
quark-mass curve
plotted as a black line with a cyan error band.

\section{$z$-expansion}
Because only relatively small kaon momenta are accessible in the lattice
calculation, the range of $q^2$ directly calculable is close to the
maximum value.  However, the experimental results are best at low
$q^2$.  We use the $z$-expansion to extend the kinematic
range~\cite{Boyd:1994tt}.  
This expansion is based on the field theoretic principles 
analyticity and crossing symmetry.
It is systematically improvable by adding more orders to the expansion.
The $q^2$ region is mapped to $z$ by
\begin{equation}
z(q^2,t_0)=\frac{\sqrt{t_+-q^2}-\sqrt{t_+-t_0}}{\sqrt{t_+-q^2}+\sqrt{t_+-t_0}},
\end{equation}
where  $t_\pm=(M_B\pm M_K)^2$.  We are allowed to choose $t_0$ as we wish
to minimize the range of $|z|$.  It
is convenient to choose  $t_0=t_+\left (1-\sqrt{1-\frac{t_-}{t_+}}\right )$.
This results in the full range of $q^2$ for this decay being mapped to
$|z| <0.16$.  The form factors are then expressed as a function of $z$:
\begin{equation}
f(q^2)=\frac{1}{B(z)\phi(z)}\sum_{k=0}^\infty a_k z^k, \nonumber
\end{equation}
where the Blaschke factor $B(z)=z(q^2,m_R^2)$
is used to account for the pole structure of the form
factor, and the outer function $\phi(z)$ is selected such 
that $\sum_{k=0}^{\infty}a_k^2\leq 1$~\cite{Arnesen:2005ez}.

\begin{figure}
\begin{center}
\begin{tabular}{c c}
\includegraphics[width=0.45\textwidth]{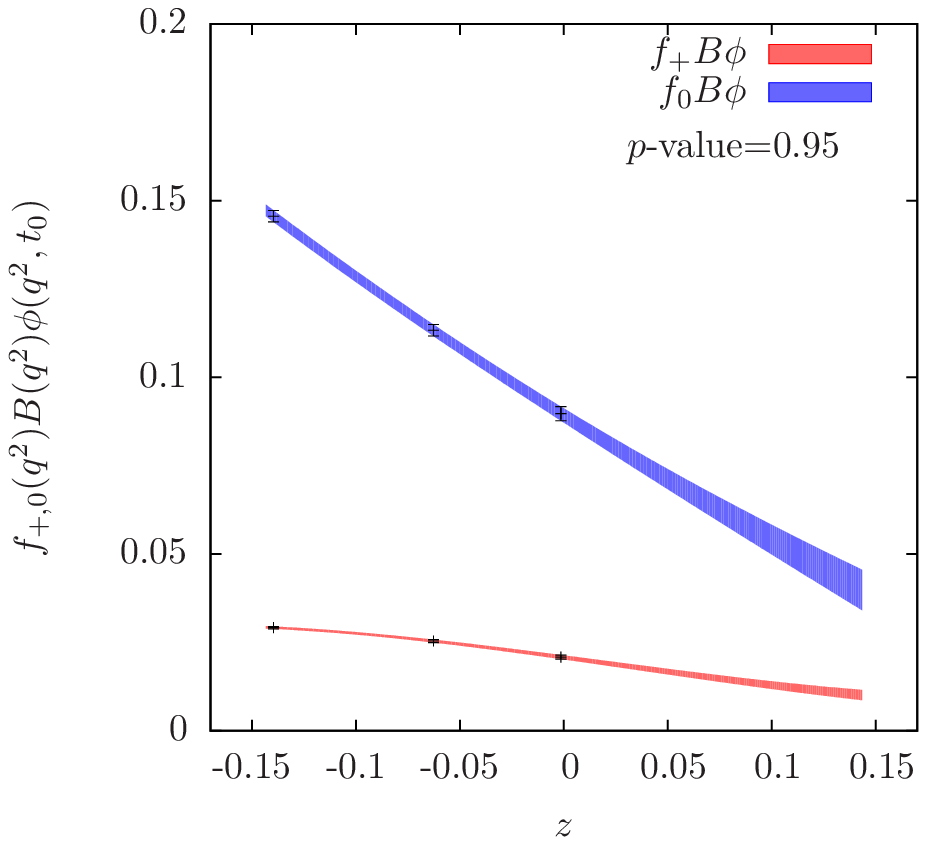}
&
\includegraphics[width=0.45\textwidth]{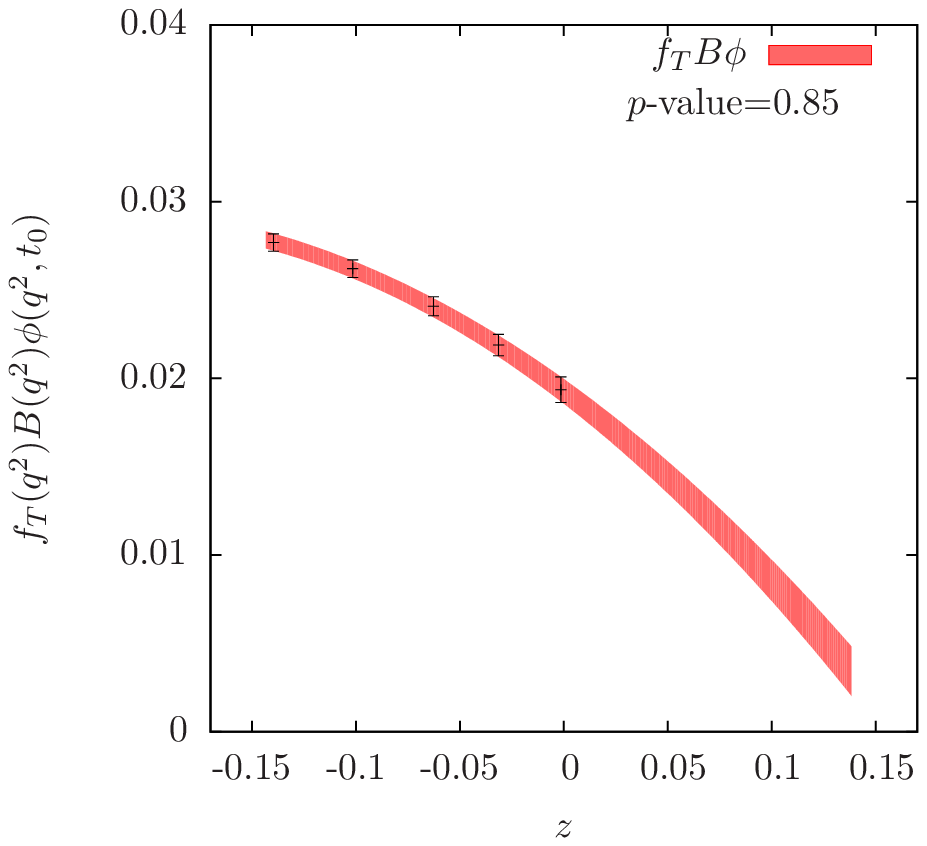}
\end{tabular}
\end{center}

\caption{Fit of form factor multiplied by Blaschke factor and outer function to
a polynomial expansion in $z$.  Only statistical errors are shown.
\label{fig:zexp}
}
\end{figure}

\begin{figure}
\begin{center}
\begin{tabular}{c c}
\includegraphics[width=0.45\textwidth]{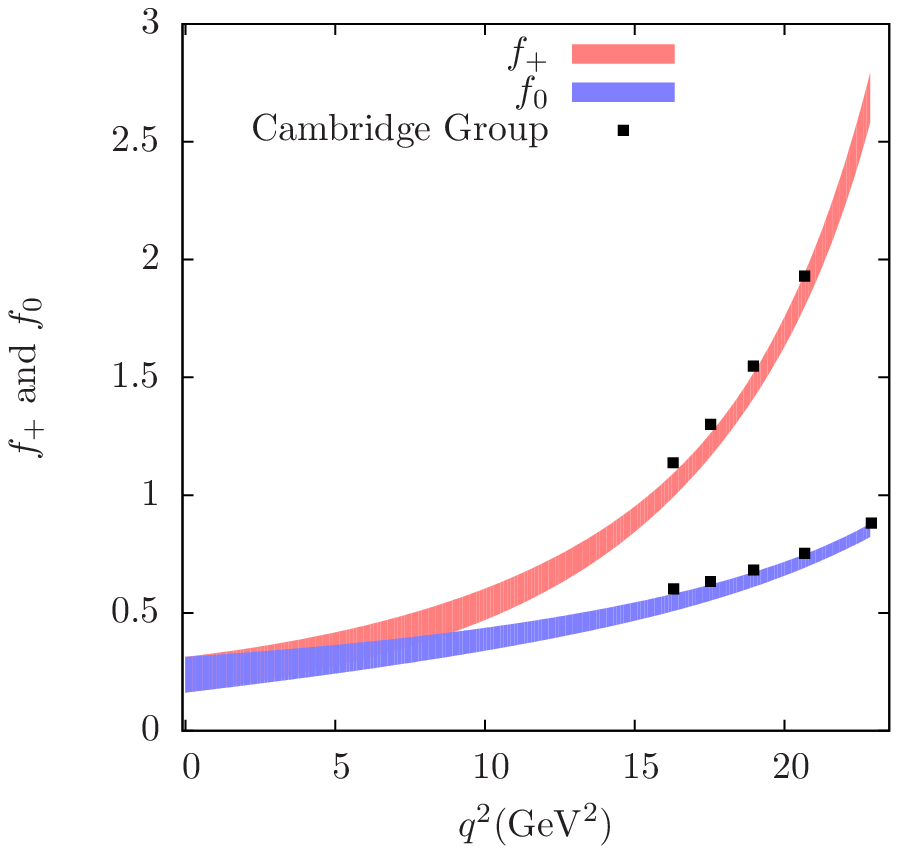}
&
\includegraphics[width=0.45\textwidth]{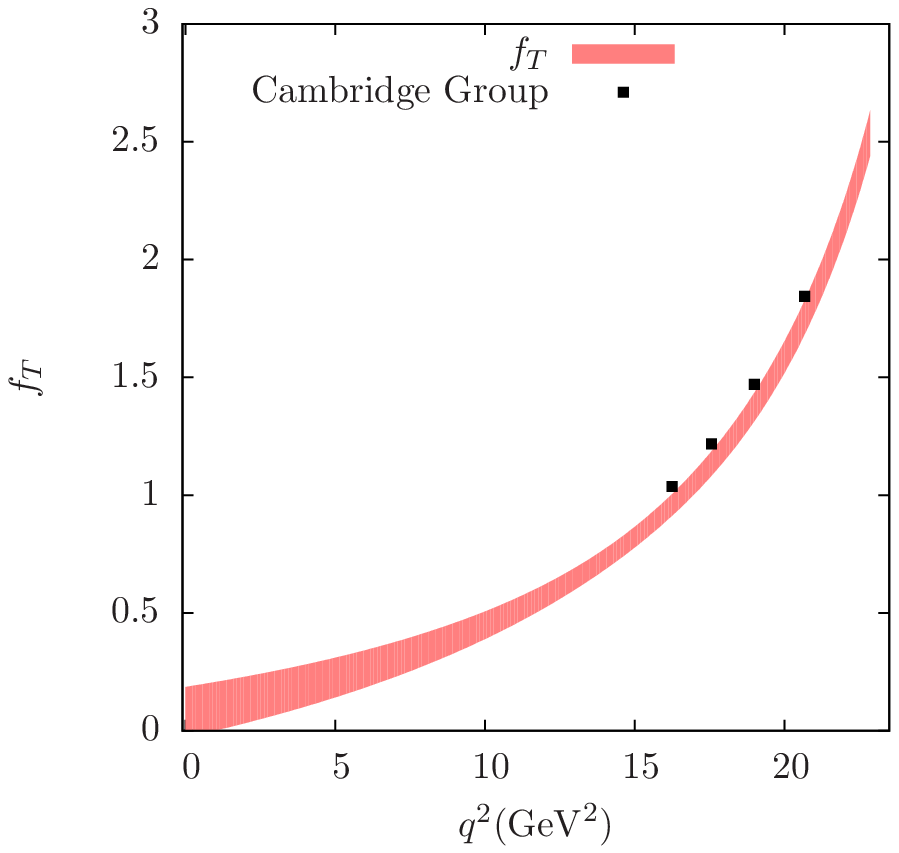}
\end{tabular}
\end{center}

\caption{Preliminary results for form factors as a function of $q^2$,
including comparison with Ref.~\cite{Liu:2011raa}.
\label{fig:zexp.sys}
}
\end{figure}

In Fig.~\ref{fig:zexp}, we show our polynomial fits
to the form factors multiplied by the Blaschke factor and outer function,
{\em i.e.}, $f(q^2)B(z) \phi(z)$.  To generate these plots, we calculate
synthetic data at selected values of $z$ in the
range accessible to our simulations based on the chiral-continuum
extrapolation of the lattice data in Figs.~\ref{fig:fpara}--
\ref{fig:ft}.  
We see that the negative $z$ region corresponds to maximum $q^2$, and that
is the region in which the lattice calculation is done.
We impose the kinematic constraint  $f_+(q^2=0)=f_0(q^2=0)$ in
the $z$-fit.
We use terms up to $z^3$ for $f_+$ and $f_0$, and $z^2$ for $f_T$.
The form factors as a function of $q^2$ are shown in
Fig.~\ref{fig:zexp.sys}.
Both statistical and systematic errors are shown,
but these results are preliminary and do 
not include the final renormalization factors or all the ensembles
that we will analyze.  

At large $q^2$, where we can
directly compute the form factors, the error is about 5\%.  As $q^2$ decreases,
the errors from statistics, chiral fit and $z$-expansion grow to
more than 30\% for $f_+$ and $f_0$ and to about 100\% for $f_T$.  One reason
that the relative errors grow at small $q^2$ is that the form factors themselves
are getting much smaller. 

\section{Conclusions}
We have presented some preliminary results
for the rare decay $B\to Kll$ form factors.
There are additional ensembles not yet included in the analysis.
We still need to include the perturbative
renormalization factors, which are expected to be close to unity
and will not change our results significantly.
Our results for $f_\parallel$, $f_\perp$ and $f_T$ 
can be used to calculate the SM $B\to K l l$ branching fraction 
or to make predictions for this process in BSM theories.
Other semileptonic decays currently
being analyzed include $B\rightarrow \pi l\nu$
and $B_s\rightarrow K l\nu$.

\section*{Acknowledgments}
Computations for this work were carried out with resources provided by
the USQCD Collaboration, the Argonne Leadership Computing Facility, and
the National Energy Research Scientific Computing Center,
which are funded by the Office of Science of the
U.S. Department of Energy; and with resources provided by the National Institute
for Computational Science, 
and the Texas Advanced Computing Center, which are funded
through the National Science Foundation's Teragrid/XSEDE Program.
This work was supported in part by the U.S. Department of Energy under Grants
No.~DE-FG02-91ER40661 (S.G., R.Z.),
No.~DE-FG02-91ER40677 (D.D., R.D.J., A.X.K.), 
No.~DE-FG02-91ER40664 (Y.L., Y.M.).
Fermilab is operated by Fermi Research Alliance, LLC, under Contract
No.~DE-AC02-07CH11359 with the United States Department of Energy.
J.A.B. is supported by the Creative Research Initiatives program 
(2012-0000241) of the NRF grant funded by the Korean government (MEST).


\begin{thebibliography}{99}
\bibitem{Aubert:2003cm} 
  B.~Aubert {\it et al.}  [BABAR Collaboration],
  Phys.\ Rev.\ Lett.\  {\bf 91}, 221802 (2003)
  [arXiv:hep-ex/0308042].

\bibitem{Abe:2001dh} 
  K.~Abe {\it et al.}  [BELLE Collaboration],
  Phys.\ Rev.\ Lett.\  {\bf 88}, 021801 (2002)
  [arXiv:hep-ex/0109026].

\bibitem{Aaltonen:2011cn} 
  T.~Aaltonen {\it et al.}  [CDF Collaboration],
  Phys.\ Rev.\ Lett.\  {\bf 106}, 161801 (2011)
  [arXiv:1101.1028 [hep-ex]].

\bibitem{Aaij:2012cq} 
  R.~Aaij {\it et al.}  [LHCb Collaboration],
  JHEP {\bf 1207}, 133 (2012)
  [arXiv:1205.3422 [hep-ex]].

\bibitem{Abada:1995fa} 
  A.~Abada {\it et al.}  [APE Collaboration],
  Phys.\ Lett.\ B {\bf 365}, 275 (1996)
  [arXiv:hep-lat/9503020].
\bibitem{DelDebbio:1997kr} 
  L.~Del Debbio {\it et al.}  [UKQCD Collaboration],
  Phys.\ Lett.\ B {\bf 416}, 392 (1998)
  [arXiv:hep-lat/9708008].
\bibitem{Becirevic:2006nm} 
  D.~Becirevic, V.~Lubicz and F.~Mescia,
  Nucl.\ Phys.\ B {\bf 769}, 31 (2007)
  [arXiv:hep-ph/0611295].
\bibitem{AlHaydari:2009zr} 
  A.~Al-Haydari {\it et al.}  [QCDSF Collaboration],
  Eur.\ Phys.\ J.\ A {\bf 43}, 107 (2010)
  [arXiv:0903.1664 [hep-lat]].
\bibitem{Liu:2011raa} 
  Z.~Liu, S.~Meinel, A.~Hart, R.~R.~Horgan, E.~H.~M\"uller and M.~Wingate,
  arXiv:1101.2726 [hep-ph].
\bibitem{Zhou:2011be} 
  R.~Zhou {\it et al.}  [Fermilab Lattice and MILC Collaborations],
  PoS LATTICE {\bf 2011}, 298 (2011)
  [arXiv:1111.0981 [hep-lat]].
\bibitem{Bernard:2001av} 
  C.~W.~Bernard 
{\it et al.} [MILC Collaboration],
  Phys.\ Rev.\ D {\bf 64}, 054506 (2001)
  [arXiv:hep-lat/0104002].
\bibitem{Aubin:2004wf} 
  C.~Aubin 
{\it et al.} [MILC Collaboration],
  Phys.\ Rev.\ D {\bf 70}, 094505 (2004)
  [arXiv:hep-lat/0402030].
\bibitem{Bazavov:2009bb} 
  A.~Bazavov 
{\it et al.},
  Rev.\ Mod.\ Phys.\  {\bf 82}, 1349 (2010)
  [arXiv:0903.3598 [hep-lat]].
\bibitem{Aubin:2007mc} 
  C.~Aubin and C.~Bernard,
  Phys.\ Rev.\ D {\bf 76}, 014002 (2007)
  [arXiv:0704.0795 [hep-lat]].
\bibitem{Bailey:2008wp} 
  J.~A.~Bailey 
{\it et al.} [Fermilab Lattice and MILC Collaborations],
  Phys.\ Rev.\ D {\bf 79}, 054507 (2009)
  [arXiv:0811.3640 [hep-lat]].
\bibitem{Isgur:1990kf} 
  N.~Isgur and M.~B.~Wise,
  Phys.\ Rev.\ D {\bf 42}, 2388 (1990).
\bibitem{Boyd:1994tt} 
  C.~G.~Boyd, B.~Grinstein and R.~F.~Lebed,
  Phys.\ Rev.\ Lett.\  {\bf 74}, 4603 (1995)
  [arXiv:hep-ph/9412324].
\bibitem{Arnesen:2005ez} 
  M.~C.~Arnesen, B.~Grinstein, I.~Z.~Rothstein and I.~W.~Stewart,
  Phys.\ Rev.\ Lett.\  {\bf 95}, 071802 (2005)
  [arXiv:hep-ph/0504209].
\end{thebibliography}
\end{document}